\documentclass[fleqn,twoside]{article}
\usepackage{espcrc2}

\usepackage{psfrag}
\usepackage{amsfonts}
\usepackage[dvips,dvipdfm]{graphicx}
\usepackage[figuresright]{rotating}

\newcommand{\AmS}{{\protect\the\textfont2
  A\kern-.1667em\lower.5ex\hbox{M}\kern-.125emS}}

\title{Four loop stochastic perturbation theory in 3d SU(3)
\vskip-3.6cm\hfill\small MIT-CTP-3419; UPRF-2003-21\vskip3.6cm}

\author{F. Di Renzo\address[PARMA]
					{Dipartimento di Fisica, Universit\`a di Parma and INFN, 
				   Gruppo Collegato di Parma, Italy},
 				A. Mantovi\addressmark[PARMA],	
 				V. Miccio\addressmark[PARMA],
        Y. Schr\"{o}der\address[Boston]
        	{Center for Theoretical Physics, MIT, Cambridge, MA 02139, USA}
        	\thanks{Partially supported by the Bruno Rossi INFN-MIT exchange program}
        	}

\begin{document}

\begin{abstract}
Dimensional reduction is a key issue in finite temperature field theory. For example, when following the QCD Free Energy from low to high scales across the critical temperature, ultrasoft degrees of freedom can be captured by a 3d SU(3) pure gauge theory. For such a theory a complete perturbative matching requires four loop computations, which we undertook by means of Numerical Stochastic Perturbation Theory. We report on the computation of the pure gauge plaquet\-te in 3d, and in particular on the extraction of the logarithmic divergence at order $g^8$, which had already been computed in the continuum.
\vspace{1pc}
\end{abstract}

\maketitle

\section*{INTRODUCTION}

Numerical Stochastic Perturbation Theory (NSPT) methods \cite{NSPT} arise within the Stochastic Quantization approach \cite{SQ}  to Quantum Field Theory and provide quite a general (and rather simple) way to perform Lattice Perturbation Theory calculations in several contexts.
In the work we are going to present in these pages, we apply them to QCD at finite temperature,
in the framework of a 3d effective theory. Lattice perturbation theory calculations are needed in order to connect non-perturbative lattice results to the continuum. Hence we set up our NSPT
simulations and compute the plaquet\-te up to $g^8$ in the 3d pure gauge theory: we reach the infinite-volume value of each coefficient of the series by extrapolating the results we get for lattices of different sizes, and in particular we find a logarithmic divergence in the last coefficient.

\section{PHYSICAL FRAMEWORK}
The topic is hot QCD, and in particular the Free Energy Density, or the pressure of the quark-gluon plasma: it represents a good observable to study the deconfinement phase transition \cite{QCDphaseTr}. 
The aim is to fill the gap between perturbative methods \cite{QCDserie} and 4d finite-temperature lattice simulations \cite{4Dsim}. 
The former have to stay in the extremely high temperature sector because of the poor convergence of the series, while the highest temperatures used in lattice simulations are about $4\div5$ times the transition temperature, because of computational resource limitations.

Dimensional reduction is a way to access the intermediate regions. One integrates out the `hard' as well as the `soft' modes to get an effective theory for the `ultrasoft' modes \cite{DimReduc}. The result is a 3d SU(3) pure gauge theory, whose contribution can then be estimated with lattice Monte Carlo techniques.
To complete the matching of 3d lattice results to continuum ones \cite{3Dsim}, one needs lattice perturbation theory computations.

\section{THE METHOD}
\subsection{The Algorithm}
According to the {\it Stochastic Quantization} approach, we sample the phase-space of the field theory randomly, according to a stochastic equation, the Langevin equation,
\begin{equation}
\frac{\partial\phi(x,t)}{\partial t}=-\frac{S[\phi]}{\partial\phi(x,t)}+\eta(x,t) \;,
\end{equation}
along a new, non-physical, stochastic time $t$. This sampling is such that the average over the noise $\eta$ leads to the Feynman-Gibbs ensemble we need for functional integration:
 
\begin{equation}
\left\langle \mathcal{O}[\phi_\eta(t)]\right\rangle_\eta 
\mathop{-\hspace{-.2cm}-\hspace{-.2cm}-\hspace{-.2cm}-\hspace{-.2cm}\longrightarrow}_{t\rightarrow\infty}
\frac{1}{Z}\hspace{-.1cm}\int\hspace{-.1cm}\mathcal{D}[\phi]\mathcal{O}[\phi]\exp\{-S[\phi]\} \;.
\end{equation}
Hence, perturbation theory is performed by using the gauge-field-version of the Langevin equation,
\begin{equation}
\partial_tU_\eta=[-i\nabla S[U_\eta]-i\eta]U_\eta \;,
\end{equation}
and by replacing the gauge fields $U_\eta$ with their perturbative expansion $\sum_kg^kU_\eta^{(k)}$ in the coupling constant $g$. Then we solve numerically the resulting system of equations {\it via} discretization of the stochastic time $t=n\tau$.

\subsection{The Code}
For this purpose we set up, from scratch, a C++ code (taking a hint from the code written for APE machines in their own language). The flexibility of this language makes it possible to set up quite a general environment of {\it classes} and {\it methods} to handle lattice-structures (both `link-like' and `site-like'). 
The code, moreover, is intended for PC-cluster usage, relying on MPI language for node-communications. In order to improve code-parallelization through spreading sub-lattices over the nodes, some tricks are used, like a wide utilization of {\it pointers} and a smart allocation of physical memory especially suited to enhancing communication rates.

\subsection{Extracting data}
In practice, we evolve our system according to a discrete-stochastic-time version of the Langevin equation, keeping all the orders (in $g$) up to the one we are interested in, and we do that with different time-steps. Moreover, choosing the time steps small enough to fall in the linear region allows us to extrapolate in a simple way to the continuum-stochastic-time case. All this, then, is repeated for different lattice sizes in order to take the infinite-volume limit. The preliminary results we present here are the result of $50$ days of running on a cluster made of 10 bi-processor Athlon MP2100 PC's.

\section{RESULTS}
\begin{figure}[t]
	\begin{center}
		\psfrag{Placchetta}[cc][cc]{$\langle 1-\mathcal{P}\rangle_{\beta^{-1}}$}
		\psfrag{Reticolo}[cc][cc][.9]{$L$ --- [lattice size]}
		\psfrag{Titolo1}[bb][bb][.9]{1-loop coefficient for the}
		\psfrag{Titolo2 }[cc][cc][.9]{plaquette mean-value $\langle 1-\mathcal{P}\rangle$}
		\psfrag{Data interpolation}[cc][cc][.9]{Data interpolation}\psfrag{Analytical result}[cc][cc][.9]{Analytical result}
		\includegraphics[scale=0.3]{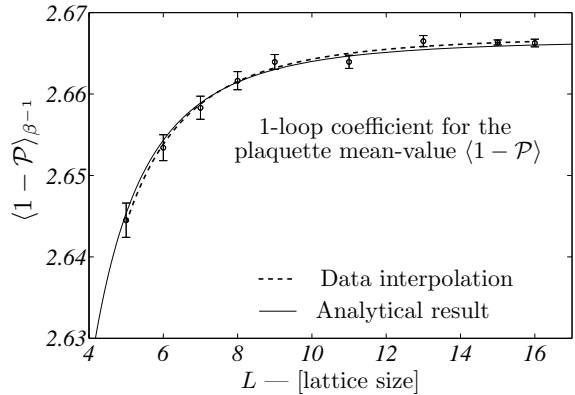}
	\end{center}
	\vspace{-1.3cm}
	\caption{Benchmark: lattice-size dependence of the $\beta^{-1}$ coefficient}
	\label{fig:g2}
	\vspace{-.5cm}
\end{figure}

We computed the weak-coupling expansion for the pure gauge plaquette $\mathcal{P} = \frac{1}{N_c}{\rm Tr}(\Pi_{\Box}{U_i})$. Our expansions are written in powers of $\beta^{-1} = \frac{g^2}{2 N_c}$. The first two coefficients of this expansion are known analytically for arbitrary dimensions \cite{first2coeff}, so we can use them as a benchmark for our code. In particular, since for the leading coefficient also the finite-size corrections are known, we can as well check each single finite-size measurement. In Fig.~\ref{fig:g2} we show the $\beta^{-1}$ coefficient {\it versus} the lattice-size. We tried to fit our numerical results (the points with their error-bars in the figure) with different forms for the finite-volume corrections, and we found that very clearly data prefer the inverse-volume correction (the dashed line in the figure).
Moreover, both the coefficient of the volume-dependence and the infinite-volume extrapolation are in good agreement with the analytical results:
\begin{eqnarray}
\left\langle 1-\mathcal{P}\right\rangle_{\beta^{-1}}^{\rm data}(L)=2.667(1) - 2.8(3) L^{-3} \;,\\
\left\langle 1-\mathcal{P}\right\rangle_{\beta^{-1}}^{\rm analytic}(L)=2.66667 - 2.66667 L^{-3} \;.
\end{eqnarray}

Also for the $\beta^{-2}$ coefficient data prefer the inverse-volume as the leading term in the `effective' finite-size correction. In fact, for different power law fits, the infinite-volume extrapolation is quite stable and, again, in good agreement with the analytical result:
\begin{eqnarray}
\left\langle 1-\mathcal{P}\right\rangle_{\beta^{-2}}^{\rm data}(L\!=\!\infty) &=& 1.95(1) \;,\\
\left\langle 1-\mathcal{P}\right\rangle_{\beta^{-2}}^{\rm analytic}(L\!=\!\infty) &=& 1.94862 \;.
\end{eqnarray}

The next coefficient, $\beta^{-3}$, is the first original result:
\begin{eqnarray}
\left\langle 1-\mathcal{P}\right\rangle_{\beta^{-3}}(L\!=\!\infty)=6.7\pm0.2\;.
\end{eqnarray}
In \cite{g6stima}, the authors estimate this coefficient rescaling the analogous 4d value (which is known, \cite{4D3thCoeff}), finding $\left\langle 1-\mathcal{P}\right\rangle_{\beta^{-3}}\approx 7.02$, which is in quite reasonable agreement with the value we found.

\begin{figure}[t]
	\begin{center}
		\psfrag{Placchetta8}[cc][cc]{$\langle 1-\mathcal{P}\rangle_{\beta^{-4}}$}
		\psfrag{Reticolo}[cc][cc][.9]{$L$ --- [lattice size]}
		\psfrag{Titolo1}[bb][bb][.9]{4-loop coefficient for the}
		\psfrag{Titolo2}[cc][cc][.9]{plaquette mean-value $\langle 1-\mathcal{P}\rangle$}
		\includegraphics[scale=0.36]{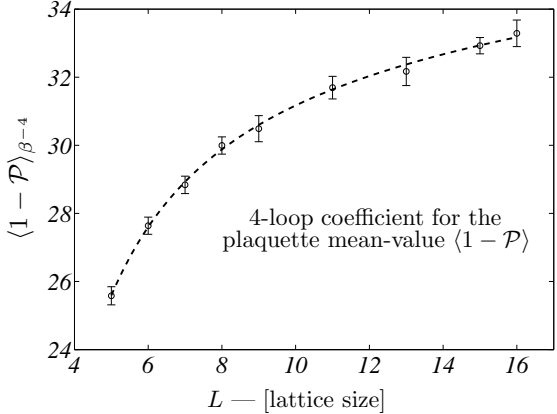}
	\end{center}
	\vspace{-1.1cm}
	\caption{The logarithmic divergence: lattice-size dependence of the $\beta^{-4}$ coefficient}
	\label{fig:g8}
	\vspace{-.4cm}
\end{figure}

As for the $\beta^{-4}$ coefficient, we expect that the lattice reflects in the volume-dependence the logarithmic divergence one finds in continuum perturbation theory \cite{LogContinuum}. Indeed we found (see Fig.~\ref{fig:g8}) that our data-fit clearly improves if we add a logarithmic term in the interpolation law. 
Writing the $\beta^{-4}$ term as 
\begin{equation}
\left\langle 1-\mathcal{P}\right\rangle_{\beta^{-4}}(L)=\ell\;\ln(L^3)+c_0+{\textstyle\sum_kc_kL^{-k}},
\end{equation}
our fits result in the following estimates:
\begin{eqnarray}
		\ell^{\rm data} = 0.9\pm0.4 \;,\quad\quad\quad c_0^{\rm data} = 23\pm5 \;,
\end{eqnarray}
where the usual difficulties in fitting a logarithmic term leads to a quite big error for the $\ell$ coefficient.

Furthermore, the coefficients of these two logarithms (the lattice one, of the lattice-volume, and the continuum one, of the cut-off of the dimensional-reduced theory) must be the same.
Hence the $\ell$ coefficient we extrapolate with our simulations can be seen as an additional (and, it turns out, positive) check for the reliability of our method. In fact, our estimate $\ell^{\rm data}$ is in good agreement with the continuum result found in \cite{LogContinuum}: $\ell^{\rm cont}=0.9765$.

As a next step, we can plug this analytical value $\ell^{\rm cont}$ into our fits in order to better estimate the constant coefficient $c_0$, for which we then get the following result:
\begin{eqnarray}
		\left. c_0^{\rm data} \right|_{\rm log\;fixed}= 25\pm2 \;.
\end{eqnarray}

\section*{CONCLUSIONS}
We use NSPT methods to compute the plaquette in a 3d pure gauge SU(3) theory, up to $g^8$.

Up to now, we are still collecting statistics in order to improve our measurements, both by increasing the number of finite stochastic-time simulations,
and by filling the gaps over the range of lattice-sizes probed.

Furthermore, our code is now ready to simulate an additional adjoint Higgs field coupled to the gauge sector, which comprises the effective theory for `soft' modes in the plasma. We are performing preliminary simulations for measuring both the quadratic and the quartic Higgs condensates.


\begin{thebibliography}{1}
\bibitem{NSPT}
F. Di Renzo, G. Marchesini and E. Onofri, P. Marenzoni, Nucl. Phys. B426 (1994) 675
\bibitem{SQ}
G.Parisi and Wu Yongshi, Sci. Sinica 24 (1981) 35
\bibitem{QCDphaseTr}
A. Papa, Nucl. Phys. B 478 (1996) 335; B. Beinlich, F. Karsch, E. Laermann and A. Peikert, Eur. Phys. J. C 6 (1999) 133
\bibitem{QCDserie}
C. Zhai and B. Kastening, Phys. Rev. D 52 (1995) 7232
\bibitem{4Dsim}
G. Boyd et al, Nucl. Phys. B 469 (1996) 419; F. Karsch, E. Laermann, A. Peikert, Phys. Lett. B 478 (2000) 447
\bibitem{DimReduc}
E. Braaten and A. Nieto, Phys. Rev. D 53 (1996) 3421
\bibitem{3Dsim}
K. Kajantie, M. Laine, K. Rummukainen, Y. Schr\"{o}der, Phys. Rev. Lett. 86 (2001) 10
\bibitem{first2coeff}
U. Heller and F. Karsch, Nucl. Phys. B251 [FS13] (1985) 254
\bibitem{g6stima}
F. Karsch, M. L\"{u}tgemeier, A. Patk\`os and J. Rank, Phys. Lett. B390, 275 (1997)
\bibitem{4D3thCoeff}
B. All\'es, M. Campostrini, A. Feo and H. Panagopoulos, Phys. Lett. B324 (1994) 433

\bibitem{LogContinuum}
K. Kajantie, M. Laine, K. Rummukainen, Y. Schr\"{o}der, Phys. Rev. D67, 105008 (2003)
\end{thebibliography}
\end{document}